\documentclass[aps,11pt,final,notitlepage,oneside,onecolumn,nobibnotes,nofootinbib,preprintnumbers,superscriptaddress,showpacs,centertags,showkeys,amsmath,amssymb]{revtex4}
\usepackage{graphicx} 
\usepackage{dcolumn} 

\def\phi{\varphi}

\begin{document}

\title{Exotic strange multibaryon states searches with $\Lambda$- hyperon and
$K^0_s$- meson systems in p+A collisions at momentum 10 GeV/c }

\author{P.Zh.Aslanyan}
\affiliation{Joint Institute for Nuclear Research, Dubna, Russia
 and Yerevan State of University}
  \email{paslanian@jinr.ru}

\begin{abstract}
Review for exotic strange multibaryon states  were obseved in the
effective mass spectra of: 1)$\Lambda \pi^+$, $\Lambda \pi^-$,
$\Lambda p$, $\Lambda p p$, and $\Lambda K^0_S$, $K^0_S\pi^{\pm}$
and $K^0_Sp$ subsystems.The invariant mass of $\Lambda \pi^+$ and
$K^0_S\pi^{\pm}$ spectra has observed well known $\Sigma^{*+}$(1385)
and $K^{*\pm}$(892) resonances. The width of $\Sigma^{*-}(1385)$ for
p+A reaction is two time larger than that presented in PDG. The
cross section of $\Xi^-\to \Lambda \pi^-$ is   7-8 times larger than
expected geometrical cross section in p+propane interaction. A few
events detected on the photographs of the propane bubble chamber,
were interpreted as S=-2 light and heavy $H^{0,+}$ dibaryons.
\end{abstract}


 \maketitle

\section{Introduction\label{sec:intro}}

There are a few  actual problems of nuclear and particle physics
which are concerning for this report. These are following goals:
in-medium modification of hadrons, the origin of hadron masses, the
restoration of chiral symmetry, the confinement of quarks in
hadrons, the properties of cold dense baryonic matter and
non-perturbative QCD, strange baryons in medium, $\Lambda$ yields,
the structure of neutron stars. Already back in 1977
Jaffe\cite{Jaffe}, using the bag model in which confined colored
quarks and gluons interact as in perturbative QCD, suggested the
existence of multi-quark states, glueballs and hybrids, but until
now none is established.Recently, the existence of discrete nuclear
bound states of $\overline{K}^0$p has been predicted with
phenomenological Kaonic Nuclear Cluster (KNC) model which is based
on the experimental information on the $\overline{K}^0$N scattering
lengths, kaonic hydrogen atom and the $\Lambda^*(1405)$
resonance\cite{knc,suzuki}.Experimental efforts for S=+1 $\Theta^+$
pentaquark have been  motivated from  report \cite{dec} where
studied antidecuplet baryons by using the chiral soliton (Skyrme)
models.

Searches for exotic strange multibaryon states with $\Lambda$ -
hypron and $K^0_S$ -meson systems were published in reports
\cite{lp}- \cite{IUTP07}.

\section{Experiment}

The full experimental information of more than 700000 stereo
photographs or $10^6$ p+propane inelastic interactions are used to
select the events with $V^0$ strange particles\cite{v0}. The masses
of the observed 8657-events with $\Lambda$ hyperon  4122-events with
$K_s^0$ meson  are consistent with their PDG values\cite{v0}. The
experimental total cross sections are equal to 13.3 and 4.6 mb for
$\Lambda$ and $K_s^0$ production in the p+C collisions at 10 GeV/c.
From published article  one can see that the experiment is
satisfactorily described by the FRITIOF model.The experimental
$\Lambda /\pi^+$ ratio in the pC reaction is approximately two times
larger than this ratio  from  pp reactions or from simulated pC
reactions by FRITIOF model\cite{fri1}  at the same energy \cite{v0}.

 For the fit of the resonance signals, the mass spectra were taken to have
the form\cite{lk,bg} $d\sigma(M)$/dm = BG(M)+BW(M)*PS(M), where
BG,BW and PS represent background, Breit-Wigner(BW) function and
phase space, respectively. The  background has been  obtained by
three methods. The first is a polynomial (or Legendre polynom)
method. The second method of the randomly mixing angle between
decayed particles from different experimental events was described
in \cite{mix,lk}.The third type of background has been obtained by
FRITIOF model\cite{fri1}.

 The statistical significance of resonance
peaks were calculated as NP /$\sqrt{NB}$, where NB is the number of
counts in the background under the peak and NP is the number of
counts in the peak above background.

\section{($\Lambda, \pi^+$) and ($\Lambda, \pi^- $) spectra}

\begin{figure}[ht]
\includegraphics[width=50mm,height=50mm]{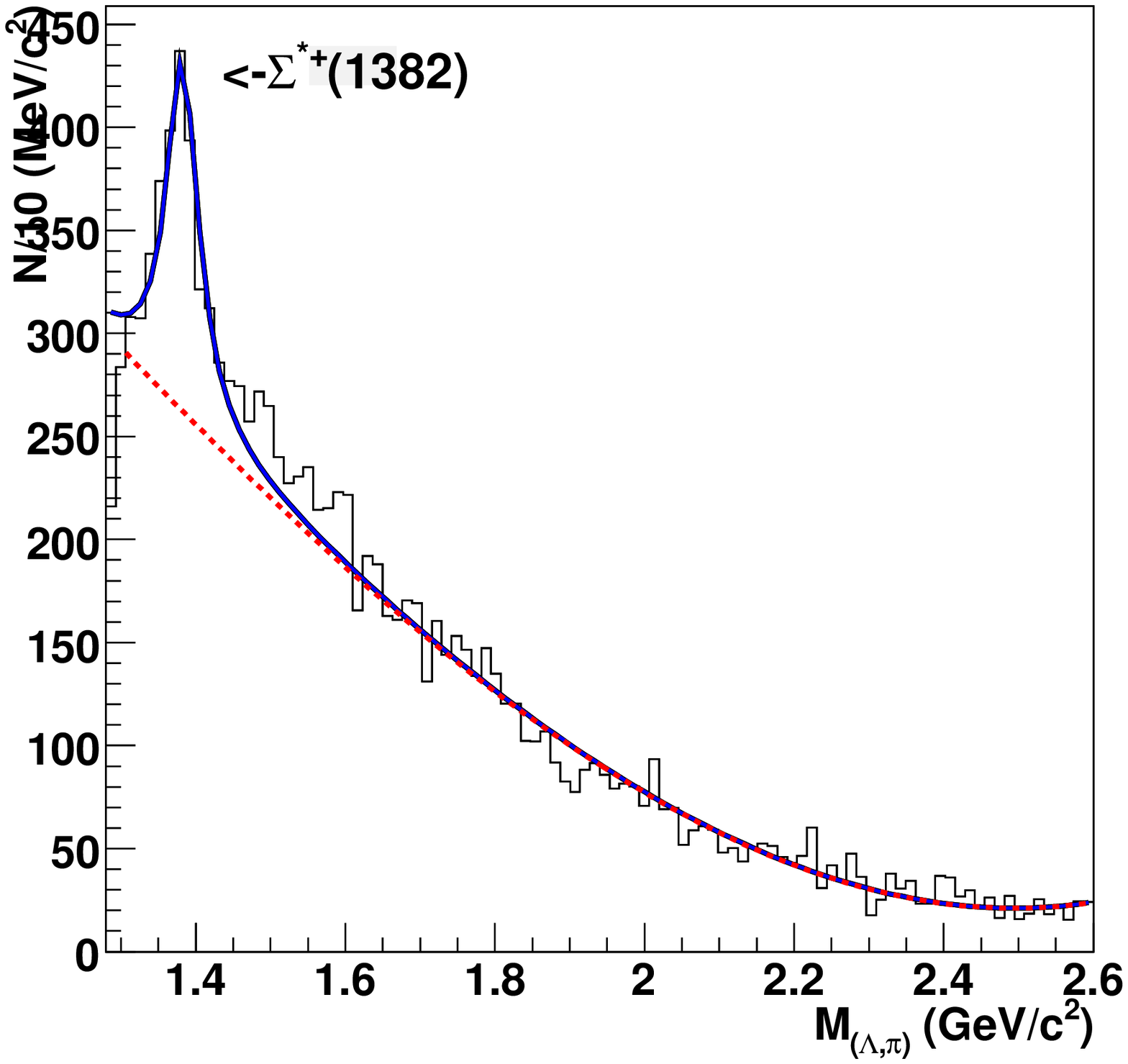}{a)}
\includegraphics[width=50mm,height=50mm]{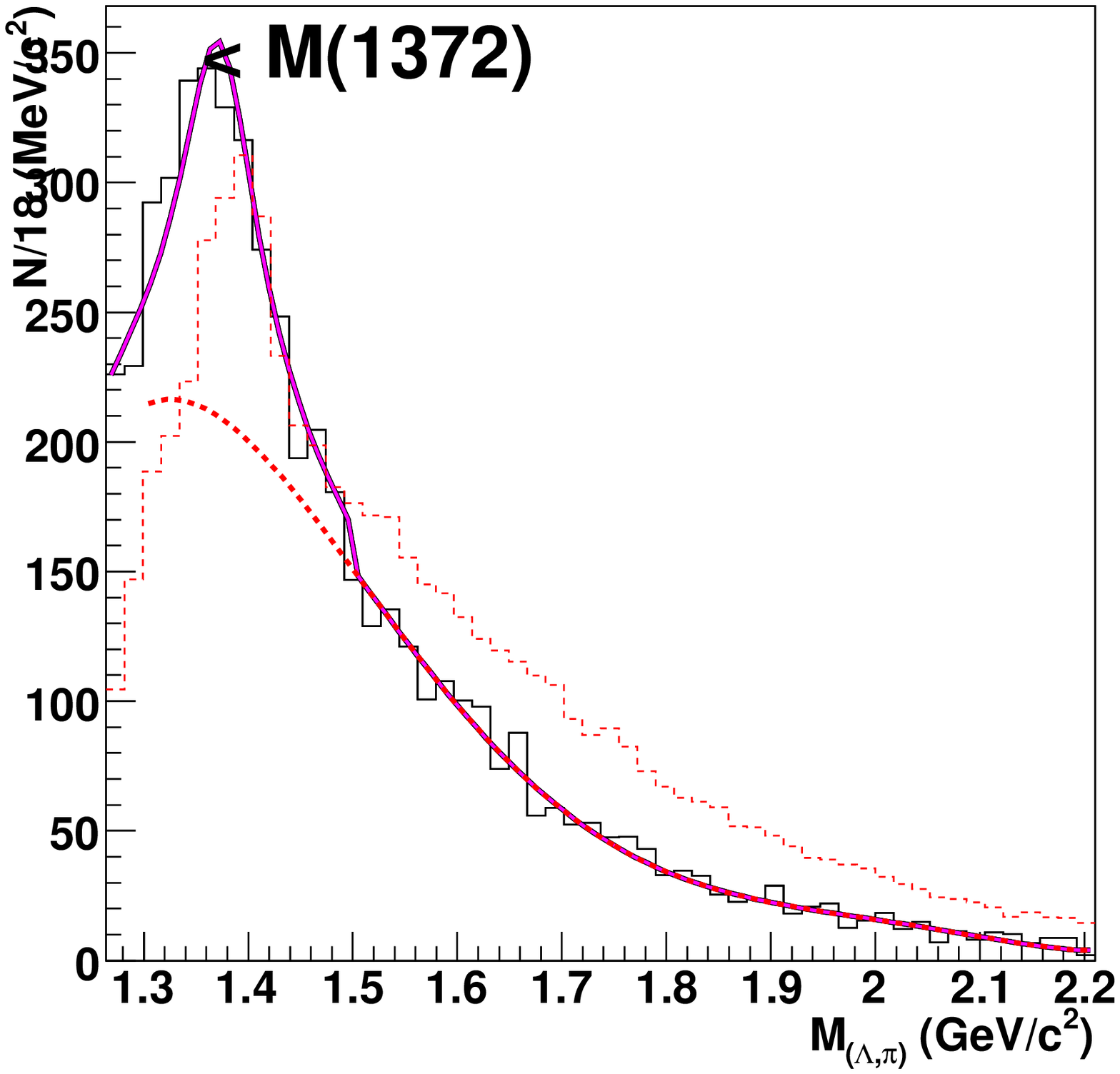}{b)}
\includegraphics[width=50mm,height=50mm]{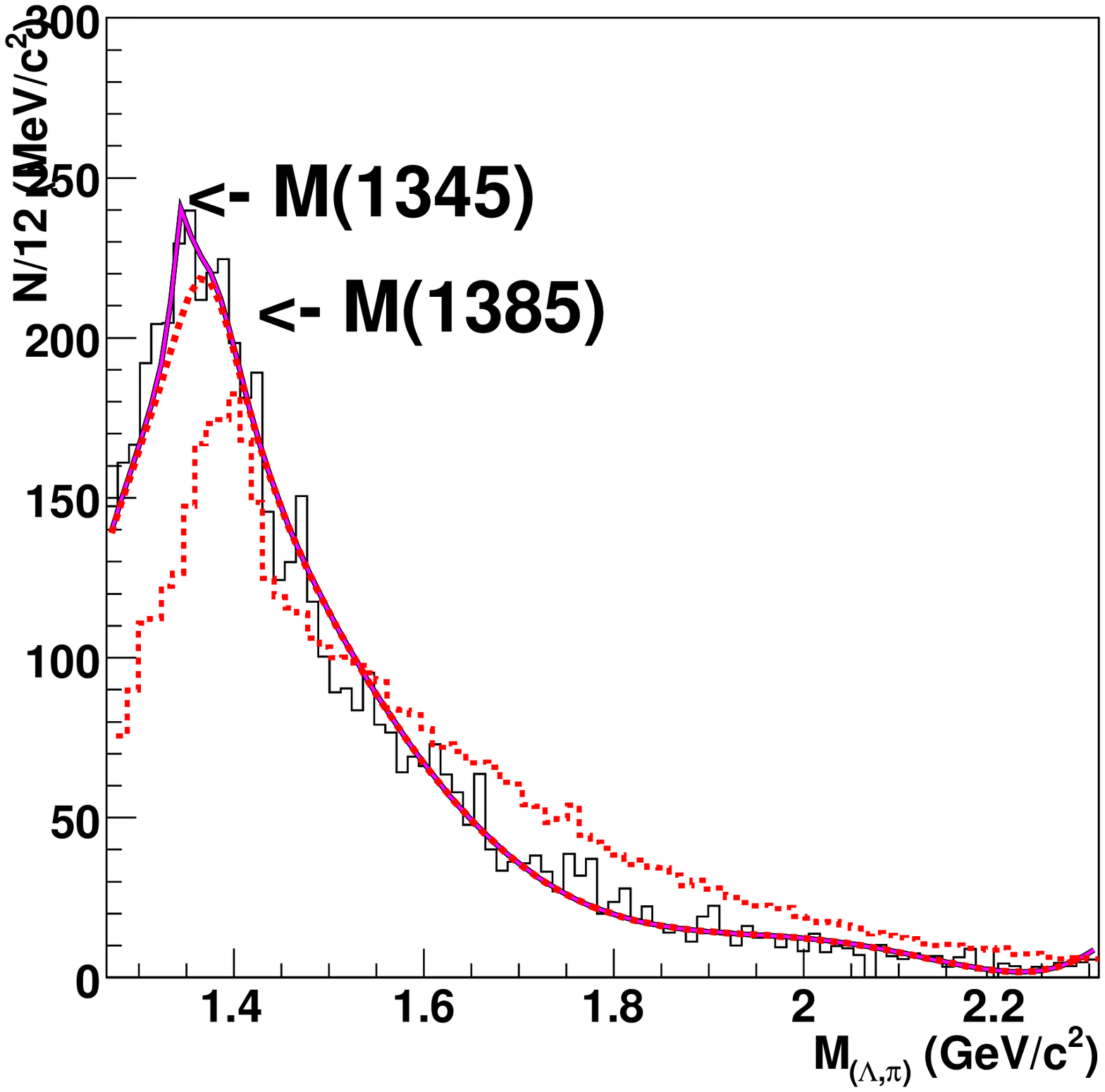}{c)}
   \caption{\label{lpi}a)The $\Lambda \pi^+$ - spectrum; b)All $\Lambda\pi^-$  comb with
  bin size of 18 MeV/$c^2$. c) $\Lambda \pi^-$spectrum with bin size of 12 MeV/$c^2$.
The simulated events by FRITIOF is the dashed histogram. The
background is the dashed curve.}
\end{figure}

The $\Lambda\pi^+$- effective  mass distribution for  all 12088
combinations with bin size of 18  MeV/$c^2$ in Fig.\ref{lpi}a
shows\cite{spin06}. The resonance with similar decay properties for
$\Sigma^{*+}(1382)\to\Lambda \pi^+$ registered which was a test for
this method (Fig. \ref{lpi}a). The decay width is equal to $\Gamma
 \approx$ 45 MeV/$c^2$. $\Delta M/M =0.7$ in range  of
$\Sigma^{*+}(1382)$ invariant mass.  The cross section of
$\Sigma^{*+}(1382)$ production (540 simulated events) can estimated
by FRITIOF model which is approximately equal to 1 mb for p+C
interaction.

The $\Lambda\pi^-$- effective  mass distribution for  all 4940
combinations with bin sizes of 18 and 12 MeV/$c^2$ in
Fig.\ref{lpi}b,c shows. The solid curve(Fig.\ref{lpi}b) is the sum
of the background (by the polynomial method ) and 1 Breit-Wigner
resonance($\chi^2/N.D.F.=39/54$). There is significant enhancement
in the mass range of 1372 MeV/$c^2$  with 11.3 S.D.,$\Gamma$ =93
MeV/$c^2$. The cross section of $\Sigma^{*-}$ production
($\approx$680 events) is equal to $\approx$ 1.3 mb at 10 GeV/c for
p+C interaction. The broadening width for $\Sigma^{*-}$ observed
$\approx$2 times larger  than PDG value. One of possible explanation
is nuclear medium effects on invariant mass spectra of hadrons
decaying in nuclei\cite{sig}.

Figure \ref{lpi}c shows effective mass distribution with  bin size
of 12 MeV/$c^2$, where there are also significant enhancements in
mass regions of 1345(3.0 S.D.) and 1480(3.2). The solid
curve(Fig.\ref{lpi}c) is the sum of the background  and 1
Breit-Wigner resonance ($\chi^2/N.D.F.=109/88$). The background
(dashed )curve is the sum of the six -order polynomial  and 1
Breit-Wigner function with parameters for identified resonance
$\Sigma^{*-}$(1385)(Fig.\ref{lpi}c). There are negligible
enhancements in mass regions of 1410, 1520 and 1600 MeV/$c^2$. The
cross section of $\Xi^-$- production ($\approx$60 events) stopped in
nuclear medium is equal to  15 $\mu$b  at 10 GeV/c for p+propane
interaction. Expected number events with $\Xi^-$ is equal 8
events(w=1/$e_{\Lambda}$ =5.3, where w is a full geometrical weight
of registred for $\Lambda$s). We observed that the experimental
production of $\Xi^-$ 7-8 times larger than the number of $\Xi^-$
events which simulated by fritiof model. Figures shows that there is
observed $\Sigma^{*-}$(1480) correlation  which is agreed with SVD2
report too\cite{svd}.

 \section{($\Lambda, p$) and ($\Lambda, p, p $) spectra}

\begin{figure}[ht]
\includegraphics[width=50mm,height=50mm]{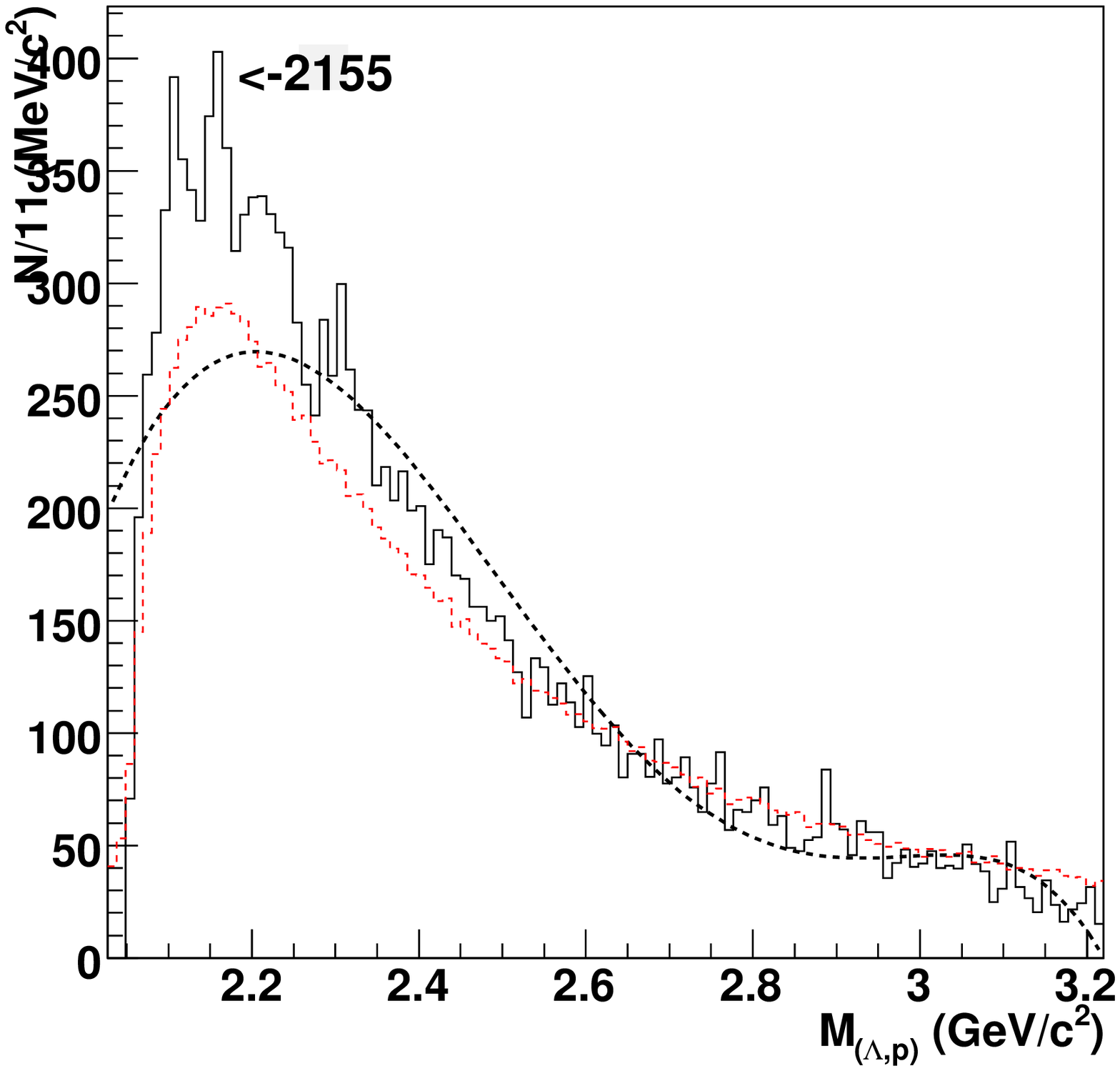}{a)}
\includegraphics[width=50mm,height=70mm]{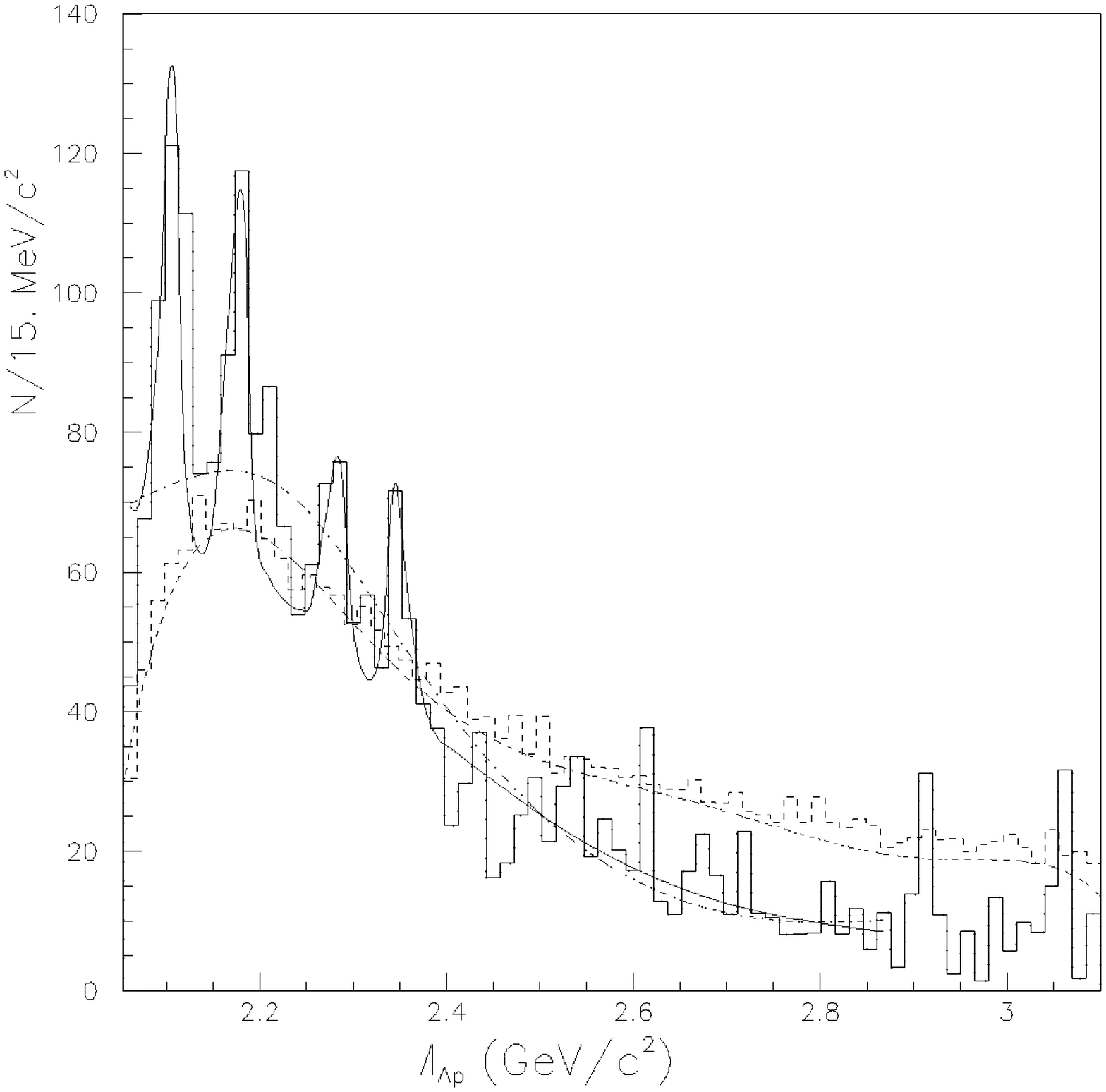}{b)}
\includegraphics[width=50mm,height=50mm]{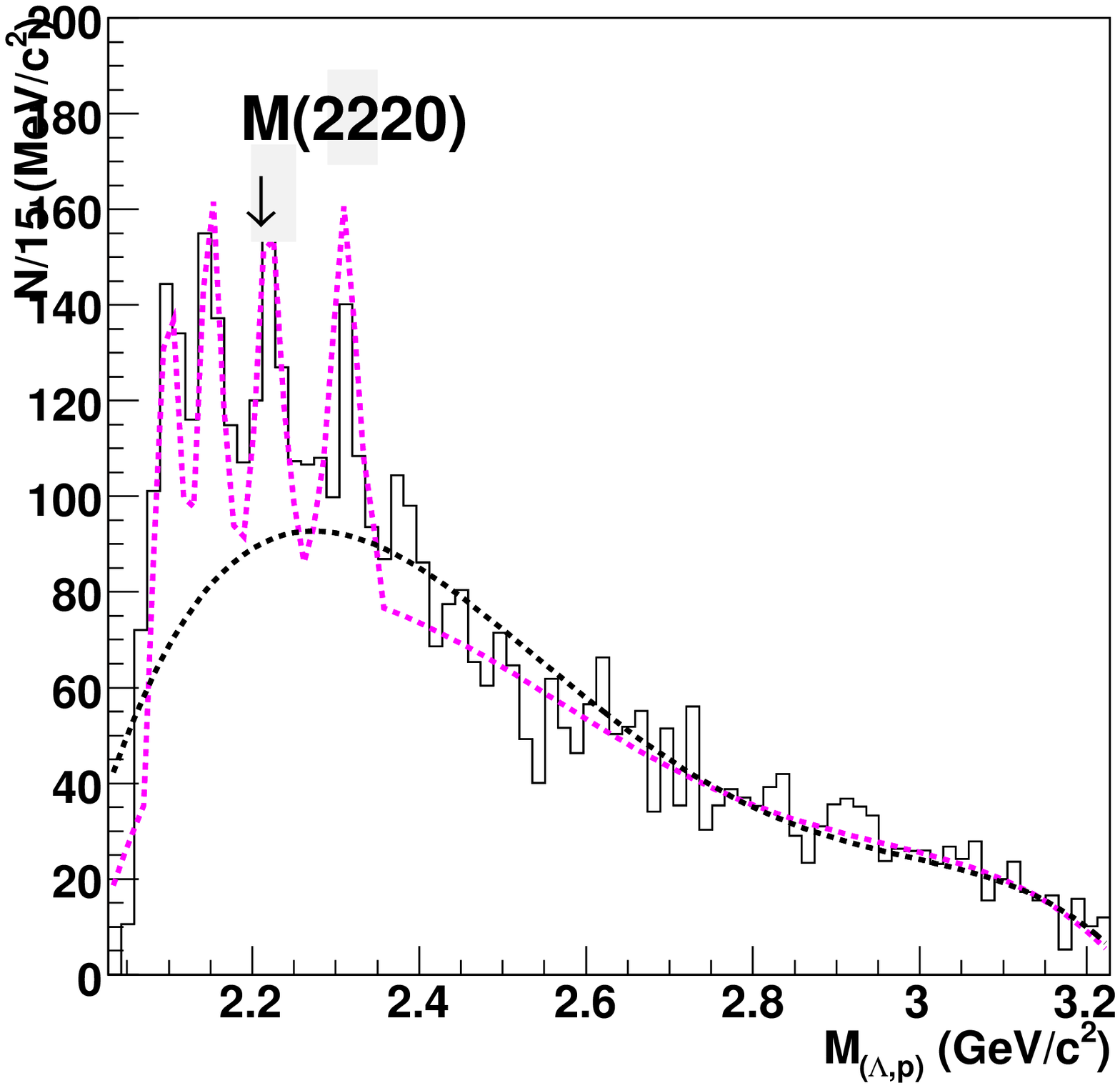}{c)}
   \caption{\label{lp}a)All comb for the $\Lambda p$ spectrum; b)$\Lambda$ p spectrum with
 identified protons in momentum range of 0.35$<P_p<$0.90 GeV/c; c) $\Lambda p$ spectrum
 with stopped protons in momentum range of 0.14$<P_p<$0.30 GeV/c.
 The dashed histogram is simulated events by FRITIOF.}
\end{figure}

 Figure  \ref{lp}a)  shows  the invariant mass  for all $\Lambda
p$ 13103 combinations with bin size of 15 MeV/$c^2 $ (\cite{lp}) .
There are  enhancements in mass regions of 2100, 2150, 2225 and 2353
MeV/$c^2$(Fig.\ref{lp}a).
 Figure \ref{lp}b  shows  the invariant mass of 2434 ($\Lambda p$)combinations
 with bin size of 15 MeV/$c^2 $(\cite{lp,spin06})for identified protons with
  momentum range of 0.350$< P_p<$ 0.900 GeV/c. There are significant enhancements
  in mass regions of 2100, 2175, 2285 and 2353 MeV/$c^2$(Fig.\ref{lp}b).Their
excess above background by the second method is 6.9, 4.9, 3.8 and
2.9 S.D., respectively. There is also a small peak in 2225( 2.2
s.d.) MeV/$c^2$ mass region.

Figure \ref{lp}c  shows the invariant mass of 4011($\Lambda
p$)combinations with bin size 15 MeV/$c^2 $ for stopped protons in
momentum range of 0.14$< P_p<$ 0.30 GeV/c.The  dashed curve is the
sum of the 8-order polynomial  and 4 Breit-Wigner curves with
$\chi^2=30/25$ from fits(Table~\ref{reslp}). A significant peak at
invariant mass 2220 MeV/$c^2$ (6.1 S.D.), $B_K$ ~ 120 MeV was
specially stressed by Professor T. Yamazaki on $\mu$CF2007, Dubna,
June-19-2007 that is conform with KNC model prediction by channel of
$K^- pp \to \Lambda $p .

\begin{figure}[ht]
\includegraphics[width=50mm,height=55mm]{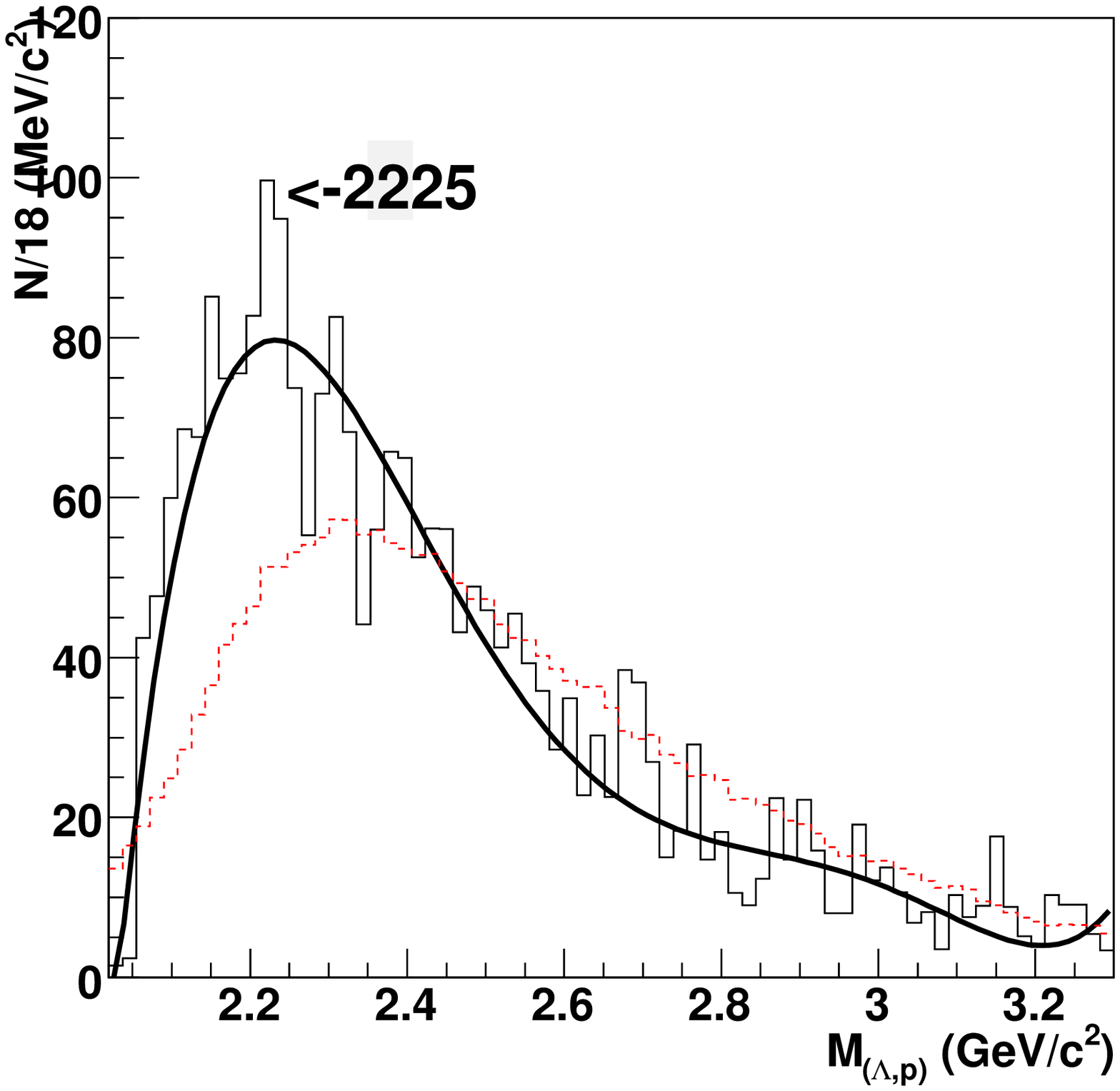}{a)}
\includegraphics[width=50mm,height=55mm]{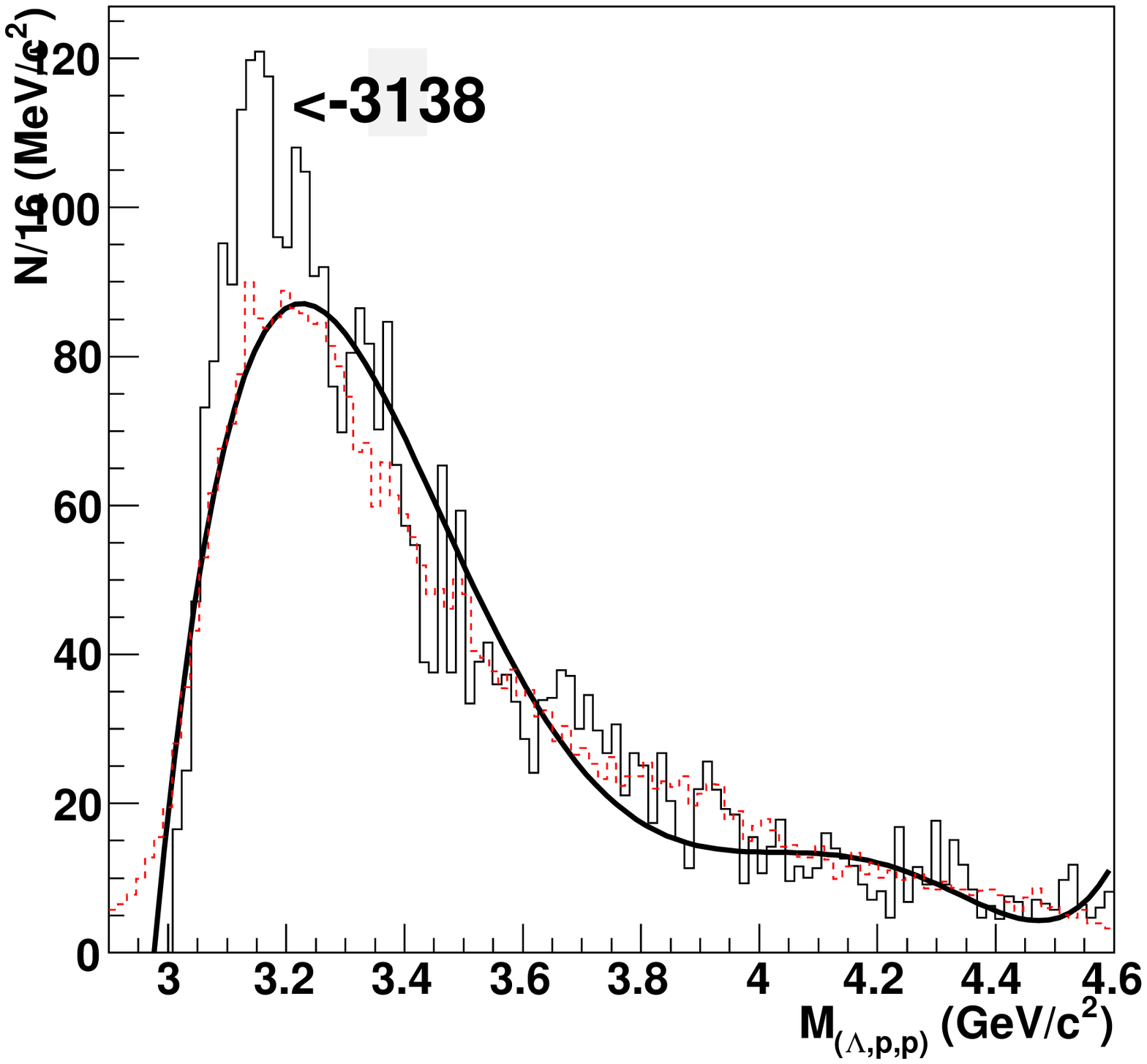}{b)}
\includegraphics[width=50mm,height=55mm]{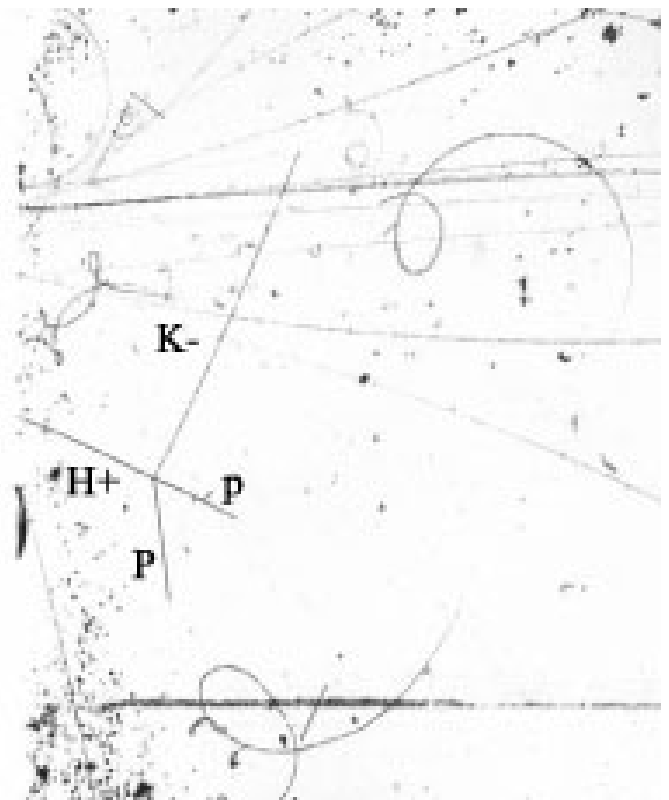}{c)}
  \caption{a)$\Lambda p$  spectrum for relativistic positive tracks
  in range of $P_p>$1.65 GeV/c;b) $\Lambda p p$ spectrum for identified
  protons; c) The weak decay for $H^+ \to K^- pp$.The dashed histogram
  is simulated events by FRITIOF. The experimental background is the solid curve.}
  \label{lpp}
\end{figure}

 The $\Lambda p$ effective mass distribution for
2025 combinations with relativistic protons over a momentum of P
$>$1.65 GeV/c is shown in Figure \ref{lpp}a . The solid curve is the
6-order polynomial function($\chi^2$/n.d.f=205/73). The background
for analysis of the experimental data are based on FRITIOF and the
polynomial method. There are significant enhancements in mass
regions of 2155(2.6 S.D.), 2225(4.7 S.D., with $\Gamma$=23
MeV/$c^2$), 2280(4.2 S.D.), 2363(3.6 S.D.) and 2650 MeV/c$^2$(3.7
S.D.). These observed peaks for combinations with relativistic
protons P $>$1.65 GeV/c agreed with peaks for combination with
identified protons and  with stopped protons.

The $\Lambda pp$   effective  mass distribution for 3401
combinations for identified protons with a momentum of  $P_p <$0.9
GeV/c is shown in Figure \ref{lpp}b\cite{spin06,jutp07}. The solid
curve is the 6-order polynomial function($\chi^2$/n.d.f=245/58,
Fig.\ref{lpp}b ). The backgrounds for analysis of the experimental
data are based on FRITIOF and the polynomial method. There is
significant enhancements in mass regions of 3138b MeV/$c^2$(6.1
S.D.) and with width 44 MeV/$c^2$. There are small enhancements in
mass regions of 3199(3.3 S.D.), 3320(5.1 S.D.), 3440(3.9 S.D) and
3652MeV/$c^2$(2.6 S.D.).  These peaks from $\Lambda p$ and $\Lambda
p p$ spectra were partly conformed  with experimental results
from FOPI(GSI), FINUDA(INFN),  OBELIX(CERN) and E471(KEK).\\

\subsection{Heavy S=-2, $H^+\to K^-pp$ dibaryon }

Stable S=-2 dibaryon state searches  are going
on\cite{H},\cite{IUTP07},\cite{spin06},\cite{sakai}. New candidates
for S=-2 $H^+$ dibaryon shows in Fig. \ref{lpp}c .The appearance of
its first part, 15.8 cm long, with a momentum of $p_{H^+} =1.2 \pm
0.12$GeV/c and average relative ionization more than $I/I_0>$2 . The
second part is due to two stopped protons. The momentum of negative
$K^-$ is equal to $0.56\pm$0.03 GeV/c($I/I_0\approx$ 1.5 ). The
kinematic threshold does not permit ($\sqrt{s}$=1.96 GeV/c)
imitating the reaction with deuteron including fermi motion. The
$H^+\to K^-pp$ hypothesis  fits the event with $\chi^2$(1V-3C)=2.6,
C.L.= 28\%, and $M_{H^+}$ =2482$\pm$48 MeV/$c^2$. There is also
possibility for fit  by hypothesis with decay channel $H^+\to
\Sigma^+\pi^-p$ which have much less probability than above
hypothesis.

\section{$K^0_sp$ - spectrum  analysis}

\subsection{$K^0_sp$ - spectrum at momentum of
$0.350\le p_p\le 0.900$ GeV/c }

  Recently there are new reports for $\Theta^+$ observation where
  statistical significance increased   for $\Theta^+ \to K^0_sp$
    and that is  equal to 7.3 S.D. from DIANA\cite{diana}
and 8.0 S.D. from SVD2\cite{svd} collaborations. The results
obtained from this experiment \cite{theta}: M $M_{\Theta}+$ =
(1540$\pm$8) MeV/$c^2$, $\Gamma$=(9.2 $\pm$1.8) MeV/$c^2$($\Gamma$
=(9.2$\pm$0.3) MeV/$c^2$, from PDG-04).

 The $K^0_sp$ effective mass distribution for 2300  combinations is shown
in Fig.~ref{kp}a)\cite{theta}. The solid curve is the sum of the
background and 4 Breit-Wigner resonance curves.The $K^0_sp$
invariant mass spectrum shows resonant structures with $M_{K_s^0
p}$=1540$\pm$8, 1613$\pm$10, 1821$\pm$11 MeV/$c^2$ and
$\Gamma_{K_s^0 p}$= 9.2$\pm$1.8, 16.1$\pm$4.1, 28.0$\pm$9.4
MeV/$c^2$. The statistical significance of these peaks has been
estimated as 5.5, 4.8 and 5.0 s.d., respectively. There are also
small peaks in mass regions of 1690( 3.6 s.d.) and 1980(3.0 s.d.)
MeV/$c^2$ .The primary total cross section for $\Theta^+{(1540)}$
 production in $p+C_3H_8$-interactions is estimated  to be $\approx~90 \mu$b.
 The experimental spectrum for $\Theta^+$ agree with the
calculated rotational spectra from the theoretical reports of D.
Akers\cite{akers}., V.H.MacGregor, A.Nambu, P.Palazzi
\cite{palazzi}, A.A. Arkhipov \cite{arkhipov}.

There were  similarly significant enhancements for the
($K^0_s,pos.tracks$) invariant mass distribution with a momentum
$p_p\ge $ 1.7 GeV/c (3500 combinations)in mass regions of 1487,
1544, 1612 and 1805 MeV/$c^2$\cite{theta}. Their excess above
background is 3.0, 3.9, 3.7 and 4.0 S.D., respectively. There is a
small peak in the mass region of 1685 MeV/$c^2$ .

\subsection{($K^0_s,pos.tracks$) - spectrum at momentum of $0.9 \le p_p\le 1.7$ GeV/c }

\begin{figure}[ht]
\includegraphics[width=50mm,height=75mm]{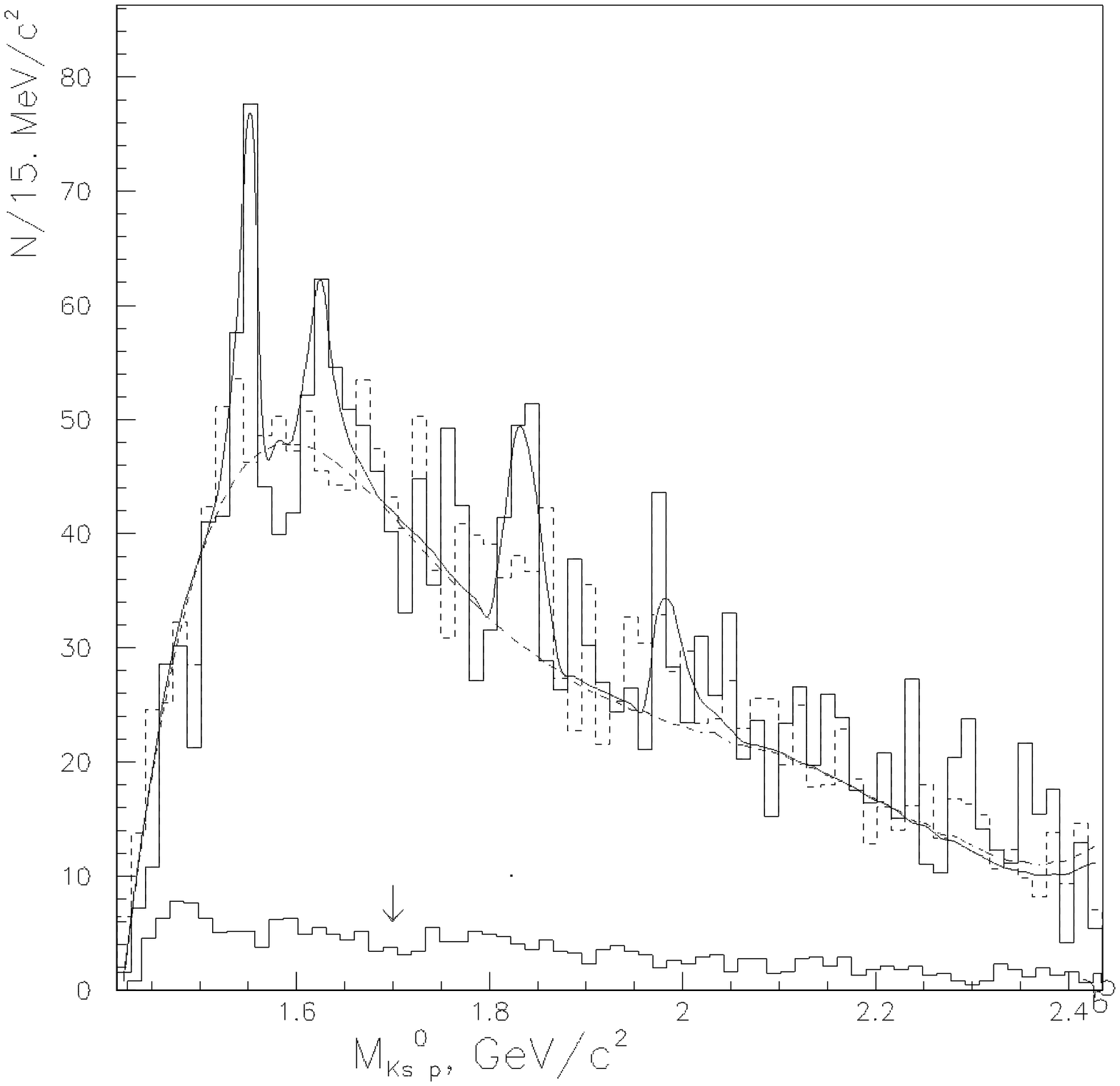}{a)}
\includegraphics[width=50mm,height=75mm]{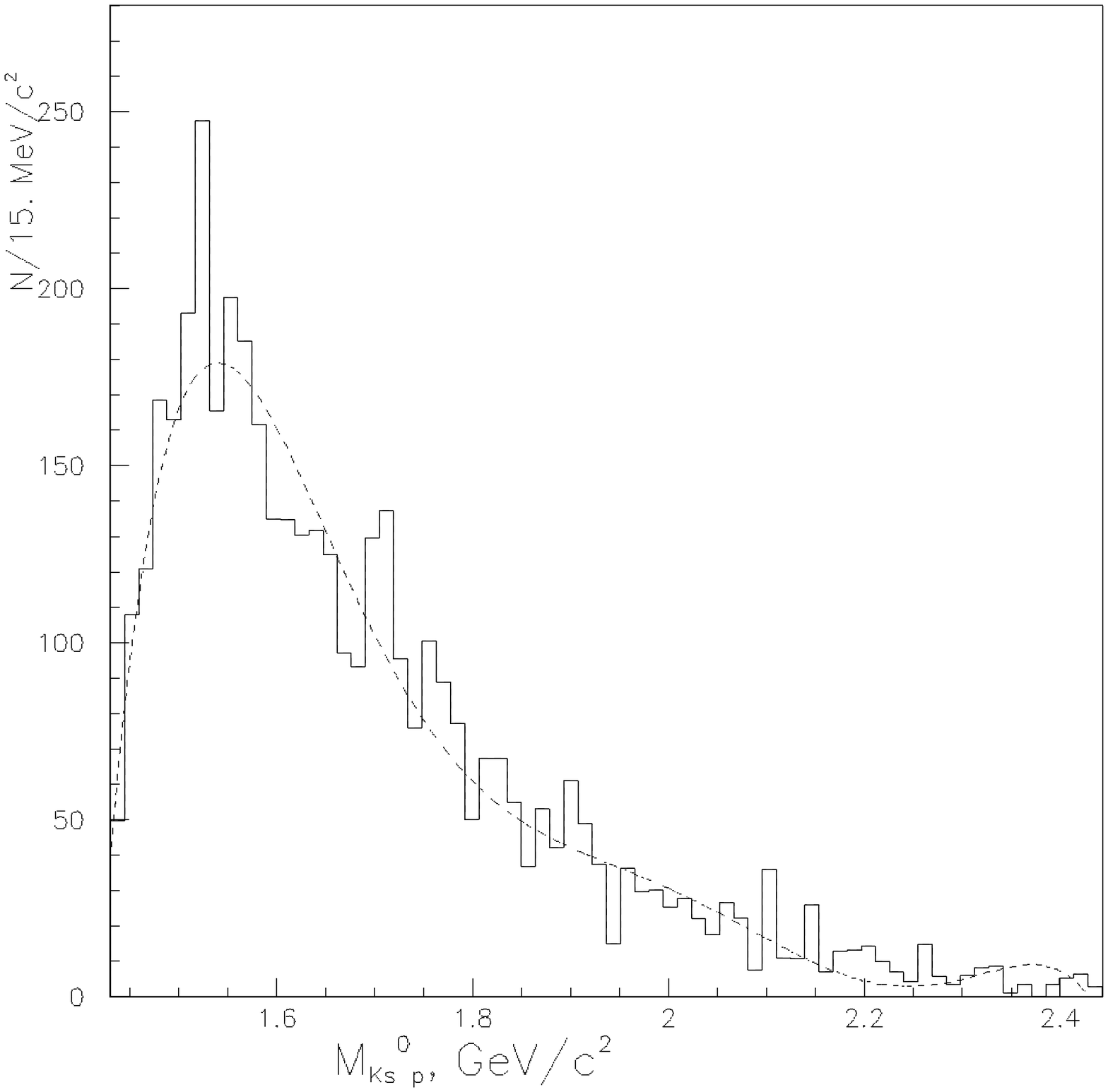}{b)}
\includegraphics[width=50mm,height=75mm]{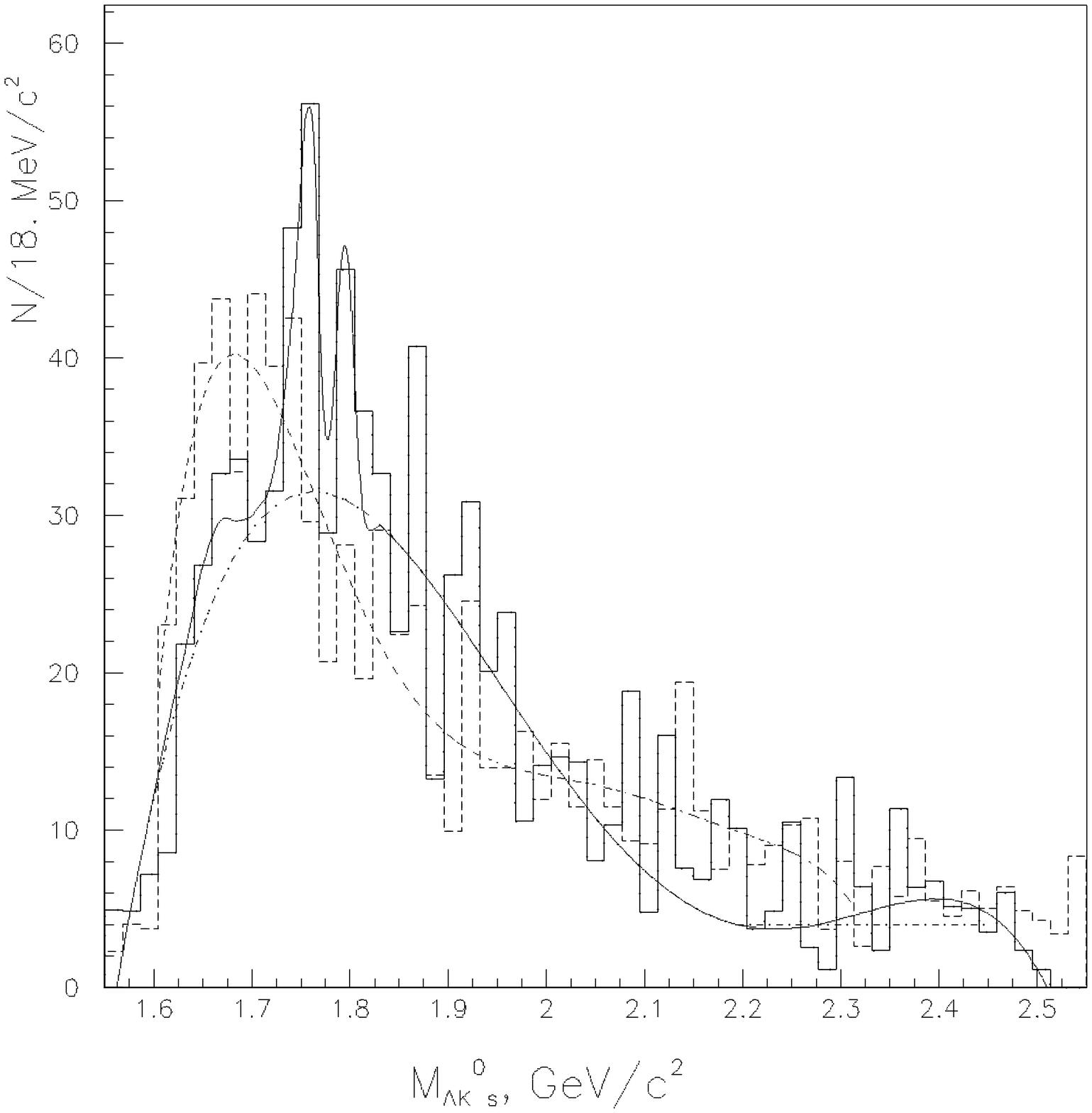}{c)}
  \caption{ \label{kp}a)$K^0_s p$  spectrum for for identified protons in range of 0.35$<P_p<$0.90
  GeV/c( $\overline{K^0}p$ comb. by FRITIOF - below histogram);
 b) ($K^0_s$, pos. relativistic tracks)  spectrum  in momentum range of 0.9$<P_p<$1.7 GeV/c;
 c) $K^0_s \Lambda$  spectrum.The dashed histogram is simulated events by
 FRITIOF.
 }
\end{figure}

 The $K^0_s,pos.track$ invariant mass spectrum shows  resonant structures with
 M = 1515 (5.3 s.d.) and 1690 MeV/$c^2$(3.8 s.d.) in Fig.~\ref{kp}b)\cite{theta}.
 No obvious structure in  mass regions of 1540,1610 and 1821 MeV/$c^2$ is seen in
  Fig.\ref{kp}b).  These observed peaks are a reflection
from resonances $\Lambda(1520)$ and $\Lambda(1700)$ in
(p$\overline{K^0}$ ) invariant mass spectrum  from
($\overline{K^0}$n ) in reactions p+p $\to K^+ (\overline{K^0}pn X$.

\section{$\Lambda K^0_s$ - spectrum  analysis}

Figure~\ref{lk}  shows  the invariant mass of 1012 ($\Lambda K^0_s$
)combinations with bin sizes 11 MeV/$c^2 $\cite{lk}.The solid curve
is the sum of the background detained by the first method and 4
Breit-Wigner  curves(Figure~\ref{lk}). A number of peculiarities
were found in the effective mass spectrum of system $\Lambda K^0_s$
in the ranges (1650-1680), (1740-1750), (1785-1805), (1835-1860) and
(1925-1950) MeV/$c^2$ in collision of a 10 GeV/c momentum with
propane. The detailed research of structure of mass spectrum has
shown, that the significant enhancements has been obtained in two
effective mass ranges 1750 MeV/$c^2$ and 1795 MeV/$c^2$. These peaks
could be interpreted as a possible candidates of two pentaquark
states: the $N^0$ with quark content udsds decaying into $\Lambda
K^0_s$ and the $\Xi^0$ quark content udssd decaying into $\Lambda
\overline{K^0_s}$. The preliminary total cross section for
$N^0(1750)$ production in p+propane interactions is estimated  to be
$\approx 30 \mu$b.

\section{$K^0_s \pi^{\pm}$ spectra analysis}

The scalar mesons have vacuum quantum numbers and are crucial for a
full understanding of the symmetry breaking mechanisms in QCD, and
presumably also for confinement\cite{kappa}.Suggestions that the
$\sigma(600)$ and $\kappa(800)$ could be glueballs have been made.

The study in \cite{k892} for vector mesons $K^{*\pm}$(892) in pp
interactions at 12 and 24 GeV/c by using data(280000 - events) from
exposure of CERN 2m hydrogen bubble chamber to p beams. Total
inclusive cross sections for $K^{*\pm}\to K^0_s\pi^{\pm}$X in pp
interactions are equal to 0.27$\pm$ 0.03 and 0.04$\pm
^{0.02}_{0.03}$ for $K^{*+}$ and $K^{*-}$ respectively.

\subsection{$K^0_s\pi^+$ - spectrum}

\begin{figure}[ht]
\includegraphics[width=50mm,height=50mm]{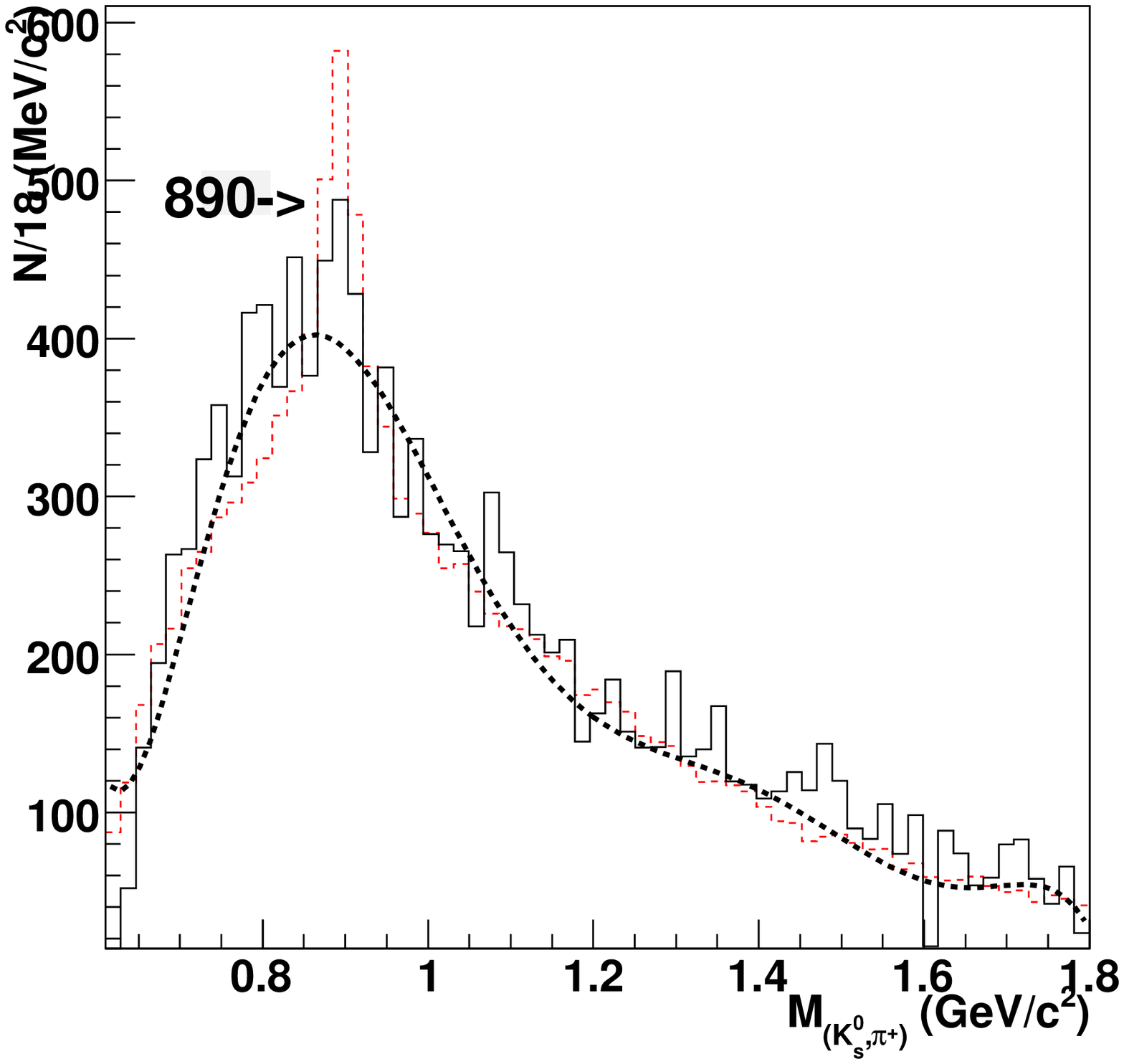}{a)}
\includegraphics[width=50mm,height=50mm]{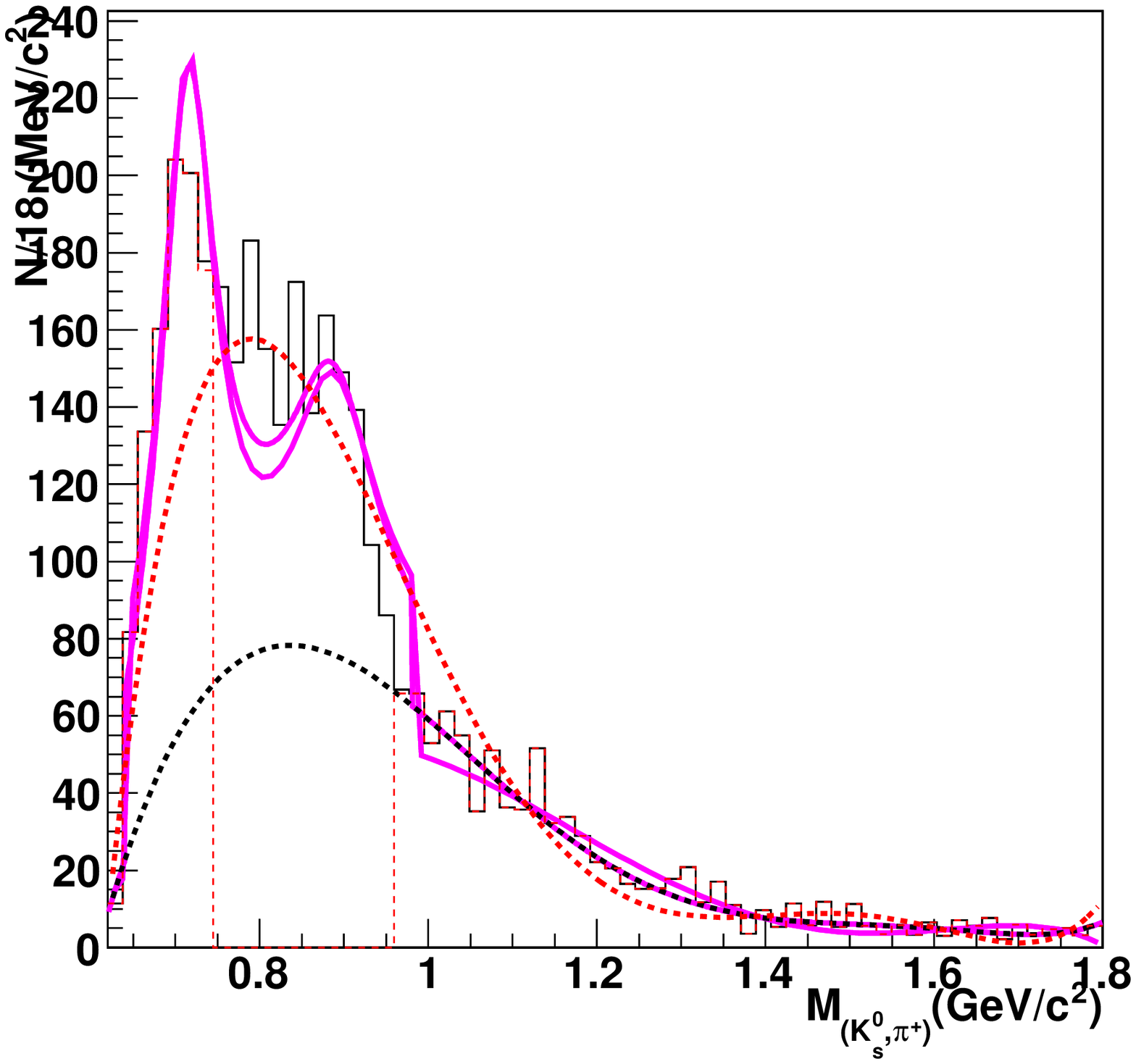}{b)}
\includegraphics[width=50mm,height=50mm]{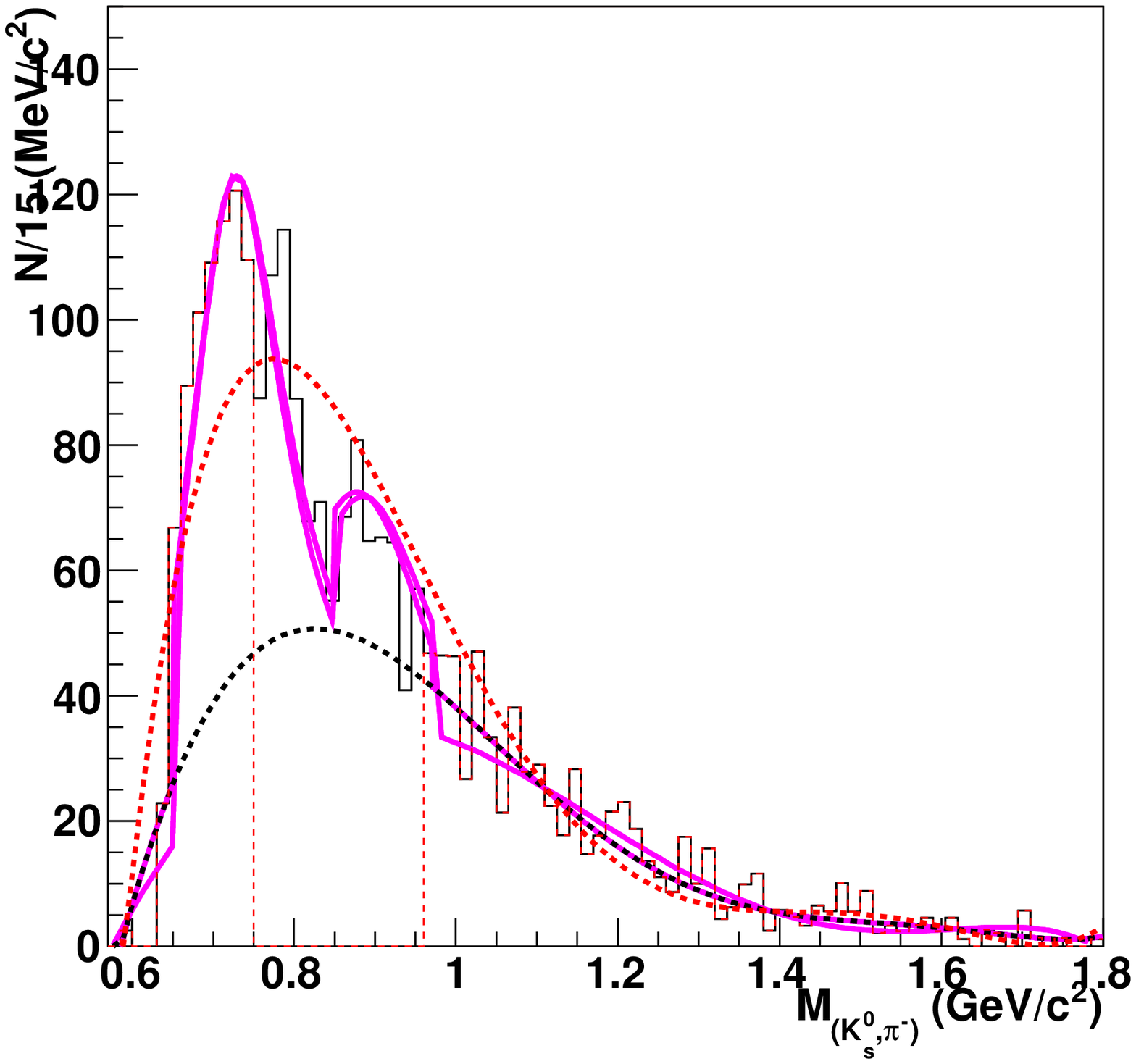}{c)}
  \caption{\label{kpipf}a)The invariant mass distribution of ( $K^0_s \pi^+$) without cuts;b)( $K^0_s \pi^+$)
  spectrum in momentum range of $P_p<$0.9 GeV/c, without background
  from comb. with protons; c)the invariant mass distribution of ( $K^0_s \pi^-$) without cuts.
  The dashed histogram is simulated events by FRITIOF. }
\end{figure}

Figure~\ref{kpipf}a)  shows  the invariant mass distribution from
all experimental 6400($K^0_s\pi^+$ )combinations with bin size of 16
MeV/$c^2$\cite{spin06},\cite{IUTP07}.   The average effective mass
resolution of $K^0_s\pi$ system  is equal to $\approx$2 \%.The
dashed curve is the background  taken in the form of a polynomial up
to the 8-th degree(Figure~\ref{kpipf}a) which agreed with background
by FRITIOF too. There are enhancements in mass regions of:
720,780,840,890 and 1060 MeV/$c^{2}$. The peak M(890)in invariant
mass spectrum is identified as well known resonance from PDG .The
preliminary interpretation of the peak in mass ranges of 1060
Mev/$c^2$ is  a reflection from well known $\Phi$ resonance by
channel of $\Phi\to K^0_s (\pi^+\pi^-)$.

The effective mass distributions of  3259($K^0_s\pi^+$ )combinations
over the momentum range of $0.05<p_{\pi^+}<0.900$ GeV/c with bin
size  18 MeV/$c^2 $ is shown in Figure~\ref{kpipf}b. Backgrounds by
FRITIOF and  polynomial methods has a similarly form. There are
enhancements in mass regions of: 720,778 and 890 MeV/$c^{2}$. The
solid curve in Figure~\ref{kpipf}b is the sum of 2BW and background
(black solid curve) taken in the form of a polynomial up to the 6-th
degree. The dashed curve(red) is the background by polynomial
without range of 0.75$< M_{K^0_s\pi} <$0.98 MeV/$c^2$ when a 1BW
function was done.

\subsection{$K^0_s\pi^-$ - spectrum}

Figure~\ref{kpipf}c)  shows  the invariant mass distribution
 of 2670 ($K^0_s\pi^-$ )combinations with bin size of 15 MeV/$c^2$\cite{spin06},\cite{IUTP07}.
 The solid curve in Figure~\ref{kpipf}c is the sum of 2BW and background
(below black solid curve) taken in the form of a polynomial up to
the 6-th degree. The dashed curve(red) is the background by
polynomial without range of 0.75$< M_{K^0_s\pi} <$0.96 MeV/$c^2$
when a 1BW function was done. There are significant enhancements in
mass regions of 720,780 and 890 MeV/$c^2$(Table~\ref{reskpi} ).The
peak 890 MeV/c$^2$ in invariant mass spectrum  is identified as well
known resonances from PDG.The preliminary total cross section for
M(720) production in p+propane interactions is larger than 30$\mu$b.

\section{Conclusion}

$\bullet{}$ Significant enhancements in invariant mass ranges of
1382 Mev/$c^2$ for $\Sigma^{*+}\to \Lambda \pi^+$   and 890
Mev/$c^2$ ($K^{*\pm}(892)\to K^0_s\pi{\pm}$ are observed
 which are test for this method and agreed with PDG .\\
$\bullet{}$A number of important peculiarities   were observed  in
the effective mass spectra for pA$\to\Lambda(K^0_s) $X reactions by
decay modes\cite{lp}-\cite{IUTP07}:$\Lambda \pi^{\pm}$, $\Lambda
p$(Table~\ref{reslp}),
$\Lambda p p$,$K^0_s\Lambda$, $K^0_s\pi{\pm}$(Table~\ref{reskpi}) and $K^0_sp$.\\
$\bullet{}$The experimental $\Lambda /\pi^+ $ ratio for average
multiplicities  in the pC reaction is approximately two
times larger than this ratio from pp reaction . \\
$\bullet{}$The width of exited $\Sigma^{*-}$(1385) is two time
larger than  PDG (preliminary result).\\
$\bullet{}$The production of stopped in medium $\Xi^-\to \Lambda
\pi^-$ is 7-8 times larger than expected geometrical cross section
for p+propane interaction (preliminary result).\\
 $\bullet{}$A few events were registered by hypothesis of
S=-2 light  and heavy $H^{0,+}$ dibaryons by weak decay
channels\cite{H},\cite{spin06}.

\begin{table}
 \caption{\label{reslp}The effective mass , width($\Gamma$) and
 S.D. for  $\Lambda p$ resonances with stopped protons in momentum range
of 0.14$< P_p<$ 0.30 GeV/c in p+ propane collisions. }
\begin{ruledtabular}
\begin{tabular}{lrrrr}  \hline
Resonance & $M_{\Lambda p}$&Experimental&&The statistical  \\
Decay & MeV/$c^2$&width $\Gamma_e$&$\Gamma$&significance \\
Mode & &MeV/$c^2$&$\approx$&S.D.\\ \hline
$\Lambda p$ &2100&36&24&5.7\\
 &2150&32&19&5.7\\
 &2220&36&23&6.1\\
  &2310&44&30&3.7\\
   &2380&46&32&3.5\\
 \hline
 \end{tabular}
\end{ruledtabular}
\end{table}

\begin{table}
\caption{\label{reskpi}The effective mass, the width($\Gamma$) and
S.D. for $K^0_s\pi^{\pm}$ resonances in p+ propane collisions.}
\begin{ruledtabular}
\begin{tabular}{lrrrr}  \hline
Resonance & $M_{K \pi}$&Experimental&&The statistical  \\
Decay & MeV/$c^2$&width $\Gamma_e$&$\Gamma$&significance \\
Mode & &MeV/$c^2$&&$S.D._{max}-S.D._{min}$\\ \hline
$K^0_s\pi^{\pm}$&890&75&50&6.0-8.2\\
  $K^0_s\pi^{\pm}$&780-800&33&10&2.5-4.2\\
$K^0_s\pi^{\pm}$&720-730&50-145&30-125&4.1-15.2\\
 \hline
 \end{tabular}
  \end{ruledtabular}
 \end{table}

\end{document}